\documentclass[aps]{revtex4}
\usepackage{graphicx}

\begin{document}

\title{Light Meson Spectra and Instanton Induced Forces}
\author{F. Brau\thanks{Chercheur I.I.S.N.} 
and C. Semay\thanks{Chercheur qualifi\'{e} F.N.R.S.}}
\affiliation{Groupe de Physique Nucl\'eaire Th\'eorique, Universit\'e
de Mons-Hainaut, B-7000 Mons, Belgium}
\date{\today}

\begin{abstract}
The spinless Salpeter equation supplemented by an instanton induced
force is used to describe the spectra of light mesons, including the
pseudoscalar ones. The coupling constants of the instanton induced
potential, as well as the quark constituent masses are not treated as
simple free parameters but are calculated from the underlying instanton
theory. Quite good results are obtained provided the quark are
considered as effective degrees of freedom with a finite size. A further
test of the model is performed by calculating the electromagnetic mass
differences between S-wave mesons.  
\end{abstract}

\maketitle

\section{Introduction}
\label{sec:intro}

Numerous papers has been devoted to the study of meson spectra in the
framework of the potential model. In most of these works, it is assumed
that the quark interaction is dominated by a linear confinement
potential, and that a supplementary short-range potential stems from the
one-gluon exchange mechanism (see for instance
Refs.~\cite{godf85,luch91}). The results obtained with these models are
generally in good agreement with the experiments but the mesons $\eta$
and $\eta'$ cannot be described without adding an appropriate flavor
mixing 
procedure with supplementary parameters.

\par On the other hand, Blask {\em et al.}\ \cite{blas90} have developed
a non-relativistic quark model which describes quite well all mesons
(including $\eta$ and $\eta'$) and baryons composed of $u$, $d$ or $s$
quarks. The long-range part of their interaction is the usual linear
confinement potential, but their short-range part is a pairing force
stemming from instanton effects. This force presents the peculiarities
to act only on quark-antiquark states with zero spin and zero angular
momentum, and to generate constituent masses for the light quarks. The
main problem of this model, and more generally of all non-relativistic
models, is that the velocity of a light quark inside a meson is not
small compared with the speed of light. This makes the interpretation of
the parameters of such models questionable \cite[p. 164]{luch91}. 

\par At present, several works have been devoted to the study of mesons
with the instanton induced forces in the framework of relativistic or
semi-relativistic models \cite{munz94,sema95,silv97}, but in all the
cases constituent masses and coupling constants of the instanton induced
forces have been considered has free parameters fitted to reproduce at
best meson spectra. Actually, these quantities can be calculated from
instanton theory. In this work, our purpose is to develop a
semi-relativistic model for meson spectra including the instanton
induced forces, but with parameters calculated, as far as possible, with
the underlying theory. We will show that such a procedure is possible
and
gives very good results provided the quarks are considered as effective
degrees of freedom with a finite size.

\par Our paper is organized as follows. In Sec.~\ref{sec:model}, the
model is constructed and the various parameters are presented, including
the quark size parameters. The numerical techniques and the fitting
procedure for the parameters are described in Sec.~\ref{sec:res}, where
the results are analyzed and a further test of the model is achieved
with the calculation of the electromagnetic mass differences. Concluding
remarks are given in Sec.~\ref{sec:concl}.  

\section{Model}
\label{sec:model}

\subsection{Spinless Salpeter equation}
\label{ssec:sse}

Our models rely on the spinless Salpeter equation. This equation is not
a covariant one, but it takes into account the relativistic kinematics.
It can be deduced from the Bethe-Salpeter equation by neglecting the
retardation effects, the mixing with negative energy states and the
spinor structure of the eigenstates. The spinless Salpeter equation has
been often used to describe meson spectra (see for instance
Refs.~\cite{godf85,silv97,fulc94}). This equation has the following form
\begin{equation}
\label{sse}
H = \sqrt{\vec p\,^2 + m_1^2} + \sqrt{\vec p\,^2 + m_2^2} + V(\vec r\,),
\end{equation}
where $V$ is the potential between the particles and where $\vec p$ is
their relative momentum. The vector $\vec p$ is the conjugate variable
of
the inter-distance $\vec r$. 

\par As usual, we assume that the isospin symmetry is not broken, that
is to say that the $u$ and $d$ quarks have the same mass. In the
following, these two quarks will be named by the symbol $n$ (for normal
or non-strange quark). 

\subsection{Funnel potential}
\label{ssec:pot}

It is now well accepted that the long range part of the interquark
interaction is dominated by the confinement. The best way to simulate
this phenomenon in a semi-relativistic equation is to use a potential
increasing linearly with the distance. With such an interaction, the
Regge trajectories of light mesons are well reproduced
\cite[p.~137]{luch91}. Moreover, lattice calculations also find that the
confinement is roughly proportional to $r=|\vec r\,|$. The short range
part is very
often a Coulomb-like interaction stemming from the one-gluon exchange
process. The idea is that, once the confinement is taken into account,
other contributions to the potential energy can be treated as residual
interactions. It is worth noting that the contribution of a constant
potential is always necessary to obtain good spectra. Finally, the
central potential considered in our models is the so-called funnel
potential \cite{luch91}
\begin{equation}
\label{funnel}
V(r) = -\frac{\kappa}{r} + a\,r + C. 
\end{equation}
Despite its simple form, this potential was the first one which is able
to reproduce the charmonium spectrum quite well \cite{luch91}. Moreover,
it gives very good results even in the light meson sector (see for
instance Ref.~\cite{fulc94}).

\par Experiment shows that vibrational and orbital excitations give much
more contributions to the meson masses than variations of spin $S$ or
total
angular momentum $J$ in a multiplet. In this case, a spinless Salpeter
equation, only supplemented by the funnel potential, can yield very
satisfactory results. Nevertheless, the situation is completely
different for the $L=0$ mesons for which the mass differences between
$S=0$ and $S=1$ states is extremely large. For these mesons, another
interaction must be taken into account in the model.

\subsection{Instanton interaction}
\label{ssec:inst}

The instanton induced interaction provides a suitable formalism to
reproduce
well the pseudoscalar spectrum. Indeed it is possible to explain the
masses of the pion and the kaon and to describe states with  flavor
mixing
as $\eta$- and $\eta'$-mesons. The form of the interaction depends on
the
quantum numbers of the state \cite{blas90}. 
\begin{itemize}
\item For $L \not= 0$ or $S \not= 0$:
\begin{equation}
\label{ins1}
V_{\text{Inst}}=0;
\end{equation}
\item For $L=S=0$ and $I=1$:
\begin{equation}
\label{ins2}
V_{\text{Inst}}=-8\, g\, \delta(\vec{r}\,);
\end{equation}
\item For $L=S=0$ and $I=1/2$:
\begin{equation}
\label{ins3}
V_{\text{Inst}}=-8\, g'\, \delta(\vec{r}\,);
\end{equation}
\item For $L=S=0$ and $I=0$:
\begin{equation}
\label{ins4}
V_{\text{Inst}}=8 
\left(
\begin{array}{cc}
g & \sqrt{2}g' \\
\sqrt{2}g' & 0
\end{array}
\right)\, \delta(\vec{r}\,),
\end{equation}
\end{itemize}
in the flavor space $(1/\sqrt{2}(|u\bar{u} \rangle+|d\bar{d}
\rangle),|s\bar{s} \rangle)$. The parameters $g$ and $g'$ are two
dimensioned coupling constants \cite{munz94} defined as 
\begin{eqnarray}
\label{ins5a}
g&=&\frac{3}{8} g_{\text{eff}}(s), \\
\label{ins5b}
g'&=&\frac{3}{8} g_{\text{eff}}(n), \\
\label{ins5c}
g_{\text{eff}}(i)&=&\left(\frac{4}{3} \pi^2\right)^2
\int_0^{\rho_{\text{c}}}
d\rho\ d_0(\rho)\, 
\rho^2\,
(m_i^0-\rho^2\, c_i),
\end{eqnarray} 
where $m^0_i$ is the current mass of the flavor $i$ and $c_i=(2/3)
\pi^2 \langle \bar{q}_i q_i \rangle$, $\langle \bar{q}_i q_i \rangle$
being the quark condensate for this flavor. The function
$d_0(\rho)$ is the instanton density as a function of the instanton size
$\rho$. For three colors and three
flavors this quantity is given by \cite{munz94},
\begin{equation}
\label{ins6}
d_0(\rho)=3.63\, 10^{-3}\left(\frac{8\pi^2}{g^2(\rho)}\right)^6\, \exp
\left(-\frac{8\pi^2}{g^2(\rho)}\right)
\end{equation}
where
\begin{equation}
\label{ins7}
\left(\frac{8\pi^2}{g^2(\rho)}\right)=9\, \ln\left(\frac{1}{\Lambda
\rho}\right)+\frac{32}{9}
\ln\left(\ln\left(\frac{1}{\Lambda \rho}\right)\right),
\end{equation}
within two-loop accuracy \cite{munz94}. The quantity $\Lambda$ is the
QCD scale parameter and $\rho_{\text{c}}$ is the maximum size of the
instanton.
This is a cutoff value for which the lnln-term in Eq.~(\ref{ins7}) is
still reasonably small compared with the ln-term. 

\par An interesting property
of the instanton induced interaction is the renormalization of quark
masses, as
it gives contributions to the constituent masses. The expression of
these contributions are given by \cite{munz94}
\begin{equation}
\label{ins8}
\Delta m_n=\frac{4}{3}\pi^2\int_0^{\rho_{\text{c}}} d\rho\ d_0(\rho)\,
(m_n^0-\rho^2\, c_n)\,
(m_s^0-\rho^2\, c_s),
\end{equation} 
and
\begin{equation}
\label{ins9}
\Delta m_s=\frac{4}{3}\pi^2\int_0^{\rho_{\text{c}}} d\rho\ d_0(\rho)\,
(m_n^0-\rho^2\, c_n)^2.
\end{equation} 
The instanton interaction is not necessarily the only source for the
constituent masses \cite{blas90}. Actually we introduce two
supplementary terms $\delta_n$ and $\delta_s$ which can be added to the
running masses. These terms are free parameters and are not dependent
on the instanton parameters. Results with vanishing and non-vanishing
$\delta_n$ and/or $\delta_s$ are given in Table \ref{tab:param}.
Finally the constituent masses in our models are given by
\begin{eqnarray}
m_n &=& m_n^0+\Delta m_n+\delta_n \\
m_s &=& m_s^0+\Delta m_s+\delta_s
\end{eqnarray} 
\par
\begin{figure}
\centering
\includegraphics*[width=10cm]{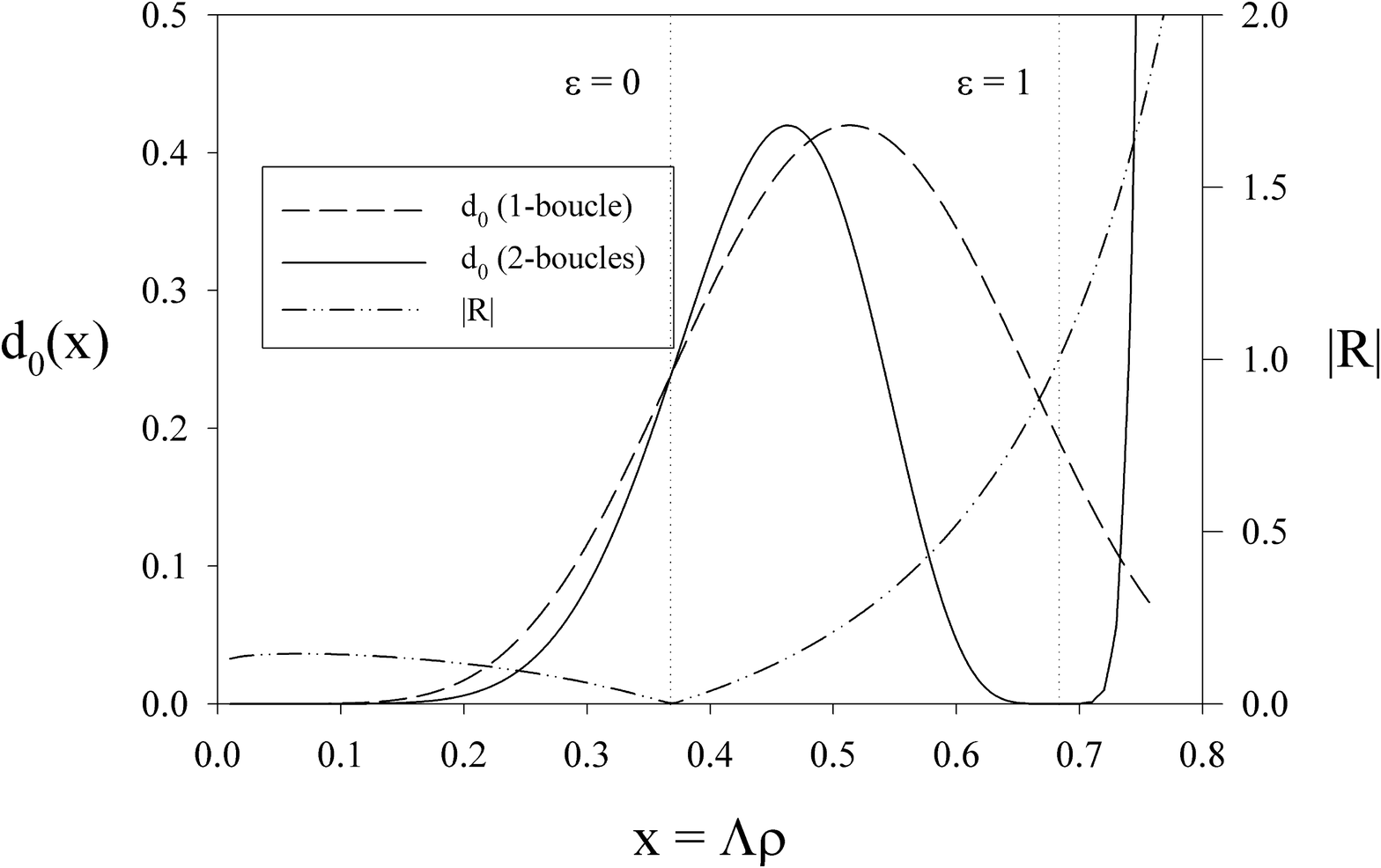}
\protect\caption{One-loop and two-loop approximations of the instanton
density (see Eq.~(\protect\ref{ins6})) as a function of the
dimensionless instanton size $x=\Lambda \rho$. The absolute value of the
ratio $R$ (see Sec.~\protect\ref{ssec:inst}), and the values of $x$
corresponding to $\epsilon=0$ and $\epsilon=1$ are indicated.}
\label{fig:d0}
\end{figure}

\begin{figure}
\centering
\includegraphics*[width=10cm]{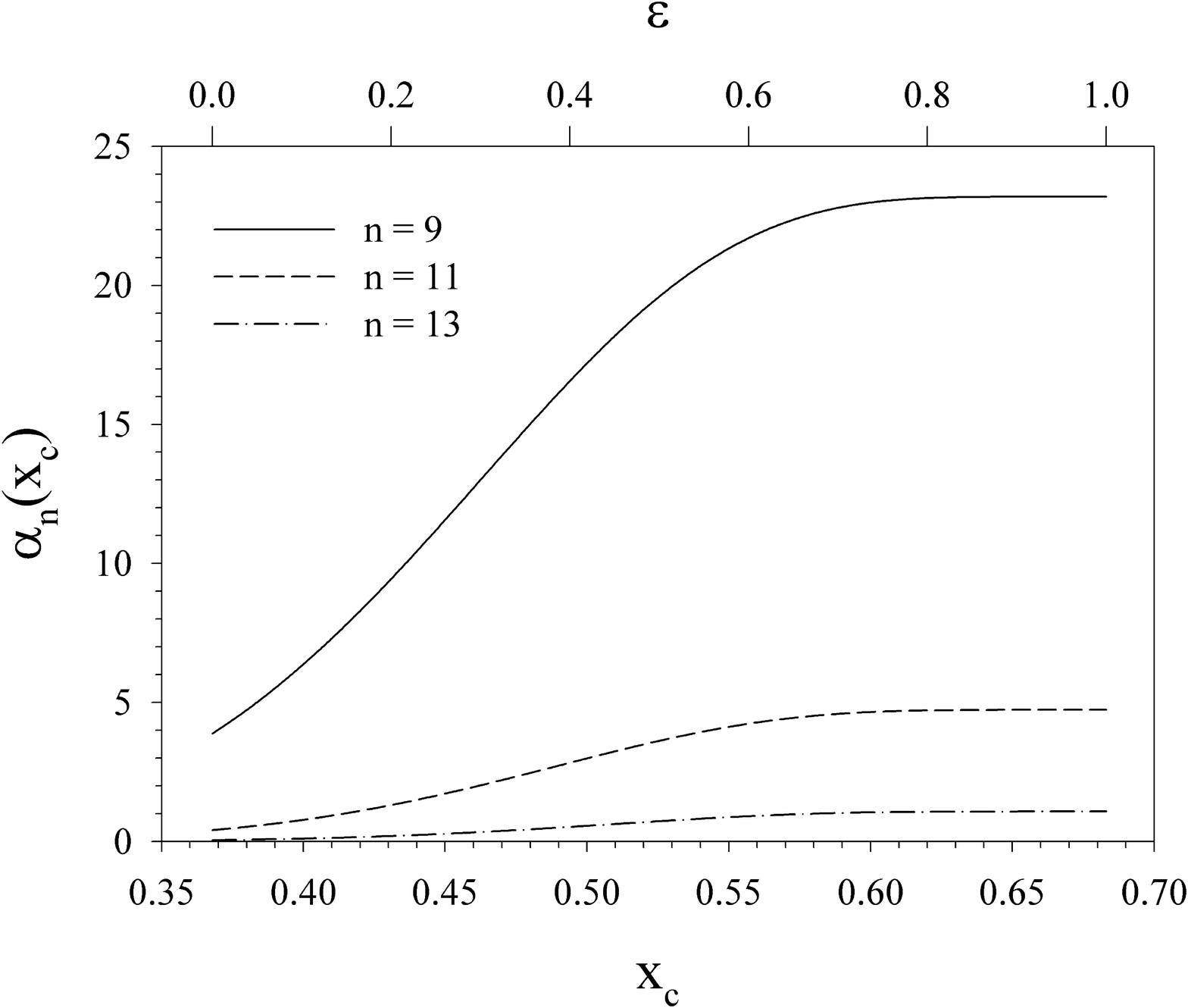}
\protect\caption{Values of the function $\alpha_n(x_{\text{c}})$ (see
Eq.~(\protect\ref{ins11})) as a function of $x_{\text{c}}$ 
and $\epsilon$ for the three
interesting values of $n$. The curves begin at $x_{\text{c}} = x_1 =
1/e$ ($\epsilon = 0$), and end at $x_{\text{c}} = x_2$ ($\epsilon =
1$).}
\label{fig:anxc}
\end{figure}

 We can rewrite expressions (\ref{ins5c}), (\ref{ins8}) and
(\ref{ins9}) in a more interesting form for
numerical calculations by setting a dimensionless instanton size
\begin{equation}
\label{ins10}
x=\Lambda \rho,
\end{equation}
and defining another dimensionless quantity 
\begin{equation}
\label{ins11}
\alpha_n(x_{\text{c}})=\int_0^{x_{\text{c}}} dx\, \left[9\,
\ln\left(\frac{1}{x}\right)+
\frac{32}{9}
\ln\left(\ln\left(\frac{1}{x}\right)\right)\right]^6\, x^n\,
\left(\ln\left(
\frac{1}{x}\right)\right)^{-32/9},
\end{equation}
where $x_{\text{c}}=\Lambda \rho_{\text{c}}$. So we obtain
\begin{eqnarray}
\label{ins12}
g&=&\frac{\delta\, \pi^2}{2}\frac{1}{\Lambda^3}\left[m^0_s\
\alpha_{11}(x_{\text{c}})-\frac{c_s}{\Lambda^2}
\ \alpha_{13}(x_{\text{c}})\right], \\
\label{ins13}
g'&=&\frac{\delta\, \pi^2}{2}\frac{1}{\Lambda^3}\left[m^0_n\
\alpha_{11}(x_{\text{c}})-\frac{c_n}{\Lambda^2}
\ \alpha_{13}(x_{\text{c}})\right], \\
\label{ins14}
\Delta m_n&=&\delta \frac{1}{\Lambda}\left[m^0_n m^0_s\
\alpha_{9}(x_{\text{c}})-\frac{(c_n
m^0_s+c_s m^0_n)}
{\Lambda^2}\ \alpha_{11}(x_{\text{c}})+\frac{c_n c_s}{\Lambda^4}\
\alpha_{13}(x_{\text{c}})\right], \\
\label{ins15}
\Delta m_s&=&\delta \frac{1}{\Lambda}\left[\left(m^0_n\right)^2\
\alpha_{9}(x_{\text{c}})-
2\frac{c_n m^0_n}{\Lambda^2}\
\alpha_{11}(x_{\text{c}})+\frac{\left(c_n\right)^2}{\Lambda^4}\
 \alpha_{13}(x_{\text{c}})\right],
\end{eqnarray}
with $\delta=3.63\, 10^{-3} \times 4\pi^2/3$. Except the quantity
$x_{\text{c}}$, all parameters involved in
Eqs.~(\ref{ins12})-(\ref{ins15}) have expected values from theoretical
and/or experimental considerations. The integration in Eq.~(\ref{ins11})
must be carried out until the ratio of the ln-term on the lnln-term in
Eq.~(\ref{ins7}) stays small . This ratio, called $R$ here, increases
with $x$ from zero at $x=x_1=1/e$ to very large values (see
Fig.~\ref{fig:d0}). At $x=x_2 \approx 0.683105$, the value of this
ration is 1. This last value corresponds to the minimum of the instanton
density (see Fig.~\ref{fig:d0}). Thus we define the parameter $\epsilon$
by
\begin{equation}
\label{ins16}
x_{\text{c}}=x_1+\epsilon(x_2-x_1) \quad \text{with} \quad \epsilon \in
[0,1].
\end{equation} 
In this work $\epsilon$ is a pure phenomenological parameter whose value
must be contained between 0 and 1. To save calculation times, we have
calculated some values of $\alpha_{n}(x_{\text{c}})$ functions for some
given values of $\epsilon$, and we use a spline algorithm to find others
values. A good accuracy can be obtained since these functions and their
derivatives are known. We have verified that the spline procedure allows
an accuracy better than $10^{-3}$ on values of quark masses and
instanton coupling constants. So, no numerical integration is necessary.
The functions $\alpha_9(x)$, $\alpha_{11}(x)$ and
$\alpha_{13}(x)$ between $x_1$ and $x_2$ are given in
Fig.~\ref{fig:anxc}.

\subsection{Effective quarks}
\label{ssec:eff}

The quark masses used in our model are the constituent masses and not
the current ones. It is then natural to suppose that a quark is not a
pure point-like particle, but an effective degree of freedom which is
dressed by the gluon and quark-antiquark pair cloud. As a correct
description of this effect is far from being obvious, we use a
phenomenological {\em Ans\"{a}tze}, as it is the case in many other
works (see for instance Refs.~\cite{godf85,silv97}). It seems natural to
consider that the probability density of a quark in the configuration
space is a peaked function around its average position. The form that we
retain is a Gaussian
function   
\begin{equation}
\label{rhoi}
\rho_i(\vec r\,) = \frac{1}{(\gamma_i\sqrt{\pi})^{3/2}} \exp(- r^2 /
\gamma_i^2).
\end{equation}
It is generally assumed that the quark size $\gamma_i$ depends on the
flavor. So, we consider two size parameters $\gamma_n$ and $\gamma_s$
for quarks $n$ and $s$ respectively. Any operator which depends on the
quark positions $\vec r_i$ and $\vec r_j$ must be replaced by an
effective one which is obtained by a double convolution of the original
bare operator with the density functions $\rho_i$ and $\rho_j$. As a
double convolution is a heavy procedure which generates very complicated
form for the convoluted potentials, we assume that the dressed
expression $\widetilde O_{ij}(\vec r\,)$ of a bare operator $O_{ij}(\vec
r\,)$, which depends only on the relative distance $\vec r = \vec r_i -
\vec r_j$ between the quarks $q_i$ and $q_j$, is given by
\begin{equation}
\label{od}
\widetilde O_{ij}(\vec r\,) = \int d\vec r\,'\,O_{ij}(\vec r\,')
\rho_{ij}
(\vec r - \vec r\,'),
\end{equation}
where $\rho_{ij}$ is also a Gaussian function of type (\ref{rhoi})
with the size parameter $\gamma_{ij}$ given by
\begin{equation}
\label{gij}
\gamma_{ij}= \sqrt{\gamma_i^2 + \gamma_j^2}.
\end{equation}
This formula is chosen because the convolution of two Gaussian
functions,
with size parameters $\gamma_i$ and $\gamma_j$ respectively, is also a
Gaussian function with a size parameter given by Eq.~(\ref{gij}).

\par After convolution with the quark density, the funnel dressed
potential has the following form
\begin{equation}
\label{vfund}
\widetilde V(r) = -\kappa \frac{\text{erf}(r/\gamma_{ij})}{r} 
+ ar \left[\frac{\gamma_{ij}\,\exp(-r^2/\gamma_{ij}^2)}{\sqrt{\pi}
\,r} + \left( 1+ \frac{\gamma_{ij}^2}{2r^2} \right)
\text{erf}(r/\gamma_{ij}) \right] + C,
\end{equation}
while the Dirac-distribution is transformed into a Gaussian function 
\begin{equation}
\label{insgauss}
\widetilde\delta(\vec r\,) = \frac{1}{(\gamma_{ij}\sqrt{\pi})^3}
\exp(-r^2/\gamma_{ij}^2).
\end{equation}
Despite this convolution, we will consider, for simplicity, that the
instanton induced forces act only on $L=0$ states.

\par The modification of the confinement potential seems very important
but, actually, only its short range part is modified. It has little
effect for a potential whose essential role is to govern the long range
dynamics. It has been verified \cite{silv97} that the convolution of the
linear potential could
be neglected without changing sensibly the results, but we have
nevertheless used the form~(\ref{vfund}) to be consistent. 

\par The Coulomb part of the interaction is transformed into a potential
with a finite value at origin. This can be seen as a mean to simulate
asymptotic freedom. It has also been noticed that the use of an error
function remove the singularity of the spin-spin interaction, when this
correction is taken into account \cite[p.~162]{luch91}.

\par The introduction of a quark size is only necessary, from a
mathematical point of view, to avoid collapse of the eigenvalues due to
the presence of a Dirac-distribution in the instanton induced
interaction. The definition of the size
parameter $\gamma_{ij}$ is not obvious in the case of the non-diagonal
term of the instanton induced interaction (\ref{ins4}). This matrix
element mixes $| n\bar n \rangle$ and $| s\bar s \rangle$ flavor states.
So, what definition choose for $\gamma_{ij}$, that we will note
$\gamma_{\text{mf}}$ in this case? A first possibility is to take
\begin{equation}
\label{gmf1}
\gamma_{\text{mf}} = \sqrt{\gamma_n^2 + \gamma_s^2},
\end{equation}
as in the case of $| \bar s n \rangle$ mesons. But we can also choose
\begin{equation}
\label{gmf2}
\gamma_{\text{mf}} = \sqrt{2\gamma_n\gamma_s},
\end{equation}
for instance. Other choices are possible, but we only take into account
these two definitions in the following. 

\section{Numerical results}
\label{sec:res}

\subsection{Numerical techniques}
\label{ssec:tech}

The eigenvalues and the eigenvectors of our Hamiltonian are obtained in
expanding trial states in a harmonic oscillator basis $|nlm\rangle$. In
such a basis, matrix elements of the potential are given by
\begin{equation}
\langle n'lm|V(r)|nlm\rangle=\sum_{p=l}^{l+n+n'} B(n',l,n,l,p)I_p,
\end{equation}
with
\begin{equation}
I_p=\frac{2}{\Gamma(p+3/2)}\int_0^{\infty} dx\ x^{2p+2} \exp(-x^2)
V(bx). 
\end{equation}   
The quantities $I_p$ are the Talmi integrals, while the coefficients
$B(n',l,n,l,p)$ are geometric factors \cite{brod67} which can be
calculated once for all. The parameter $b$ is the oscillator length
which fixes the scale of the basis states. With this parameter a
dimensionless length $x=b/r$ can be defined. In our model all the Talmi
integrals for the potential part of the Hamiltonian are given by
analytical expressions
\cite{silv97}. The matrix elements of the kinetics part can be
calculated with numerical integrations, but we prefer to use another
technique, much accurate and less time consuming. This method is
described in Ref.~\cite{fulc94}. We just give here the main hits of the
procedure. We calculate the matrix elements of the operator $\vec p\,^2
+ m^2$ in the oscillator basis (this is an analytical calculation). Let
us assume that $P$ is the corresponding matrix, $D$ the diagonal matrix
of the eigenvalues of $P$, and $U$ the transformation matrix
($D=U^{-1}\,P\,U)$. Then the matrix elements of the operator $\sqrt{\vec
p\,^2 + m^2}$, are contained in the matrix $U\,D^{1/2}\,U^{-1}$. This
procedure is about six times faster than a numerical integration and
gives a better accuracy. The utilisation of this procedure and the
analiticity of the Talmi integrals lead to very short time of
calculation. A minimization with thirteen parameters lasts about 10--20
minutes on a Pentium 200 workstation. 

\par At last, it is worth noting that all the results obtained with the
technique described above have been verified with the three-dimensional
Fourier grid Hamiltonian method \cite{brau98}.

\subsection{Fitting procedure}
\label{ssec:mini}

The purpose of this work is to try to reproduce the spectrum of light
mesons with a quite simple model. Indeed we use only a central potential
supplemented by an instanton induced interaction to describe the
pseudoscalar sector. We need thirteen parameters to obtain a
satisfactory theoretical spectrum. The instanton interaction is defined
by six parameters: The current quark masses, the quark condensates for
the flavors $n$ and $s$, the QCD scale parameter $\Lambda$, and the
maximum size
$\rho_{\text{c}}$ of the instanton. Three parameters are used for the
spin-independent part of the potential: The slope of the confinement
$a$, the
strength $\kappa$ for the Coulomb-like part, and the constant $C$ which
renormalize the energy. Four supplementary parameters are introduced:
The effective size of the quarks $n$ and $s$, and two terms $\delta_n$
and $\delta_s$, which contribute to the constituent quarks masses. 

\par In our model, the quantum numbers $L$, $S$ and $I$ are good quantum
numbers. For $L=0$ states, the instanton induced interaction raises the
degeneracy between $S=0$ (pseudoscalar) and $S=1$ (vector) mesons, but
$S=1$ $n\bar n$-mesons, which differ only by isospin, have the same
mass. This is in quite good agreement with the data. In the sector $L
\not= 0$, instanton induced interaction vanishes and the potential is
spin independent. Consequently, states which differ only by $S$, $J$ and
$I$ are degenerate. In first approximation, this corresponds also quite
well to the experimental situation. 

\par To find the value of the parameters, we minimize a $\chi^2$
function based on the masses of 18 centers of gravity (c.o.g.)\ of
multiplets containing well-known mesons (see Table~\ref{tab:meson}):
\begin{equation}
\label{chi2}
\chi^2=\sum_i\left[\frac{M^{\text{th}}_i-M^{\text{exp}}_i}{\Delta
M^{\text{exp}}_i} \right]^2,
\end{equation}
where the quantity $\Delta M^{\text{exp}}_i$ is the error on the
experimental masses (it is fixed at the minimum value of 10 MeV). The
quantum number $v$ (vibrational or radial quantum number), $L$, and $S$
of these mesons are determined on an assignment made in the Particle
Data Group tables \cite{pdg}. It is worth noting that all members
of several multiplets considered here are not known. So we calculate the
center of gravity with only the known mesons; no attempt is made to
estimate the masses of the missing states. The usual procedure to
define a center of gravity is 
\begin{equation} 
\label{cog} M_{\text{c.o.g.}}=\frac{\sum_{J,I}(2I+1)(2J+1)\
M_{J,I}}{\sum_{J,I}(2I+1)(2J+1)}. 
\end{equation} 
In the following, a center of gravity will be indicated by the name of
the state of the multiplet with the higher quantum numbers $J$ and $I$.
At last, note that we do not include the mesons $a_0(980)$ and 
$f_0(980)$ in the $L=1$ multiplet since experimental considerations and
some theoretical works \cite[p.~99, 557]{pdg} suggest that they are not
$q\bar{q}$-mesons. 

\par To perform the minimization, we use the most recent version of the
MINUIT code from the CERN library \cite{jame75}

\subsection{Meson spectra}
\label{ssec:spectra}

A great number of parameter sets have been found for our model
Hamiltonian. We present here only five among the most interesting ones,
denoted from I to V. Parameters for these models are presented in
Table~\ref{tab:param}. All these models have common features. For
instance, it is always possible to obtain a quite good fit with a low
current quark mass $m^0_n$, in agreement with the bounds expected
\cite{pdg}. The strange quark mass $m^0_s$ varies more significantly
following the model, but its value is always reasonable \cite{pdg}. The
QCD scale parameter $\Lambda$ is a very sensible parameter of our model
since a small change can largely modify the quantities deduced from the
instanton theory (see Eq.~(\ref{ins12}) to (\ref{ins15})). Nevertheless,
the values found are always in agreement with usual estimations
\cite{pdg}. Values for the quark condensates are also reasonable
\cite{rein85}. There
is no direct measurement of the string tension parameter $a$, but
lattice calculations favor the value 0.20 GeV$^2$ with about 30 \%
error \cite{mich96}. Our values are in good agreement with this
prediction. 

\par The strength $\kappa$ of the Coulomb-like potential can strongly
vary from one model to another (see for instance Ref.~\cite{luch91}).
The values found in this paper are always less than 0.5, which can be
considered as an upper limit for relevant values. In model~V, we have
fixed $\kappa=0$ as in the model of Ref.~\cite{blas90}. But the $\chi^2$
obtained is not good; this is essentially due to radial excitations:
$\rho(1450)$, $\pi(1300)$ and $\phi(1680)$. We conclude that the
Coulomb-like potential must be taken into account and that instanton
induced forces cannot explain alone the short range part of the
potential. Actually, this
remark is already mentioned in the Ref.~\cite{blas90}. As we can see
from Table~\ref{tab:param}, the constant potential is always necessary
to obtain good spectra. Its origin is not simple to explain. It can be
considered as a mechanism linked to the flux-tube model of mesons
\cite{sema95}. It has also been suggested that $C\approx -2\,\sqrt{a}$
\cite[p.~190]{luch91}. Our models are not in very good agreement with
this last prediction, as we can note deviation as large as 40 \%. But we
do not consider that it is as an important drawback of the models. 

\par
\begin{figure}
\centering
\includegraphics*[width=10cm]{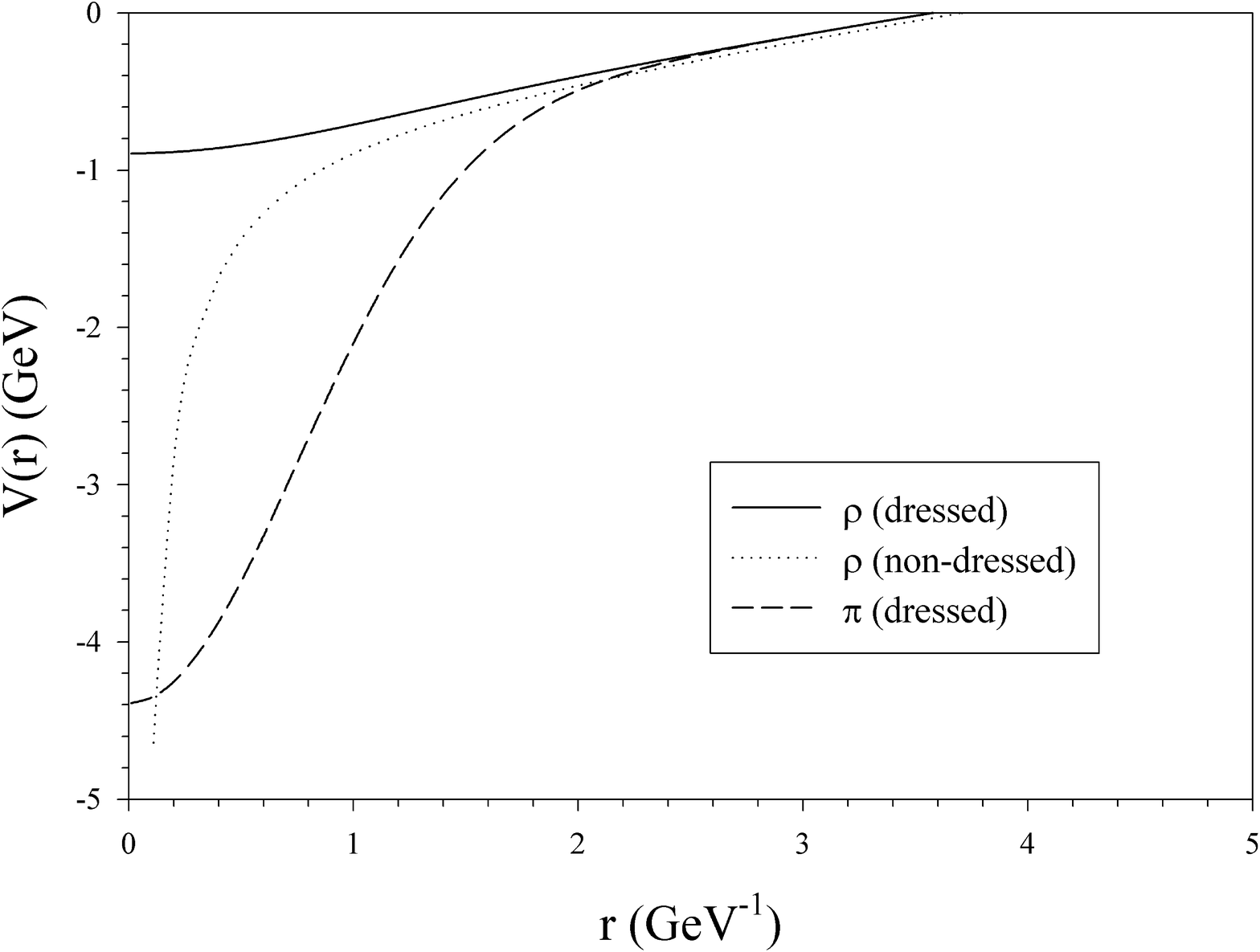}
\protect\caption{Interquark potential $V(r)$ in GeV as a function of $r$
in GeV$^{-1}$ for the $\rho$- and the $\pi$-meson. The potential for the
$\rho$-meson without taking into account the effect of the quark sizes
(non-dressed) is also presented. The corresponding potential for the
$\pi$-meson cannot be shown because the presence of a Dirac-distribution
coming from the instanton induced interaction.}
\label{fig:pot}
\end{figure}

 There is no real estimation of the quark sizes. They are purely
phenomenological parameters whose role is to take into account
relativistic effects \cite{godf85} as well as sea-quark contributions.
In all our models the size of the $n$-quarks $\gamma_n$ is nearly a
constant, around 0.7--0.8 GeV$^{-1}$. The $n$-quark size drops down to
0.6
GeV$^{-1}$ in model V without the Coulomb-like potential. This shows the
strong influence of this interaction in determining $\gamma_n$. It is
worth noting that the introduction of quark sizes modifies deeply the
structure of the potential. This is illustrated on the
Fig.~\ref{fig:pot} where the dressed and non-dressed potential for the
$\rho$-meson are presented. The $s$-quark size depends on the {\em
Ans\"{a}tze} chosen to calculate $\gamma_{\text{mf}}$. When   
$\gamma_{\text{mf}} = \sqrt{\gamma_n^2 + \gamma_s^2}$, $\gamma_s$ can
take vanishing values without generating collapse of the eigenenergies.
This is no longer true if $\gamma_{\text{mf}} =
\sqrt{2\gamma_n\gamma_s}$. The first definition of $\gamma_{\text{mf}}$
is chosen in model II and we can remark that in this case the value of
$\gamma_s$ is significantly smaller than in other models, where the
second definition is used. 

\par Table~\ref{tab:param} shows that the contributions of parameters
$\delta_n$ and $\delta_s$ can be quite large when they are not fixed in
the minimization. The origin of these terms is not clear but their
values indicate that the instanton effects cannot generate solely the
constituent masses in our model Hamiltonian. We have found sets of
parameters which give reasonable $\chi^2$ values but, in
this case, the constituent quark masses can be very small. Model~III has
actually the lowest $\chi^2$ value that we have found but the
constituent quark masses are so small that a generalization of this
model to baryon spectra appears very problematic \cite{stan}. 

\par In all sets of parameters that we have determined, the value of
$\epsilon$ is always close to zero. This is consistent with the fact
that the cutoff radius for the integration over instanton density must
be small enough in order that the lnln-term in Eq.~(\ref{ins7}) must be
reasonably small compared with the ln-term. Actually, we can fix
arbitrarily $\epsilon=0$ without spoiling the results, while values
close to unity yield bad results.

\par
\begin{figure}
\centering
\includegraphics*[width=10cm]{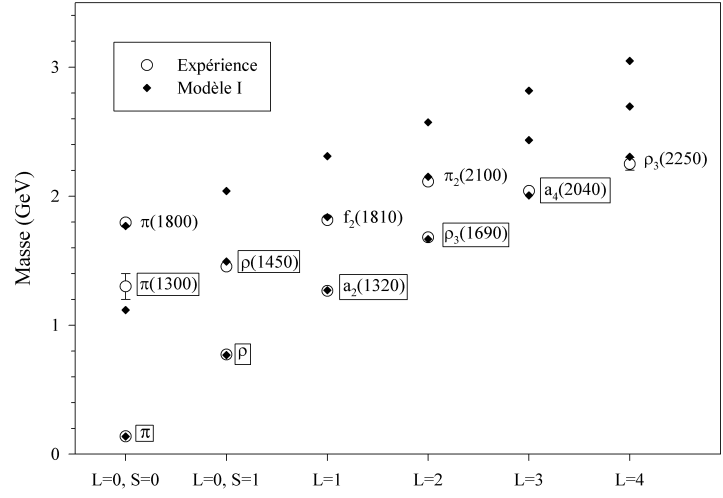}
\protect\caption{Comparison of experimental (open circles with the
corresponding error bars) and calculated (black diamonds) spectra of $n
\bar n$-mesons for the model~I. Framed names indicate centers of gravity of
multiplets used to fix the parameters.}
\label{fig:spec1}
\end{figure}

\begin{figure}
\centering
\includegraphics*[width=10cm]{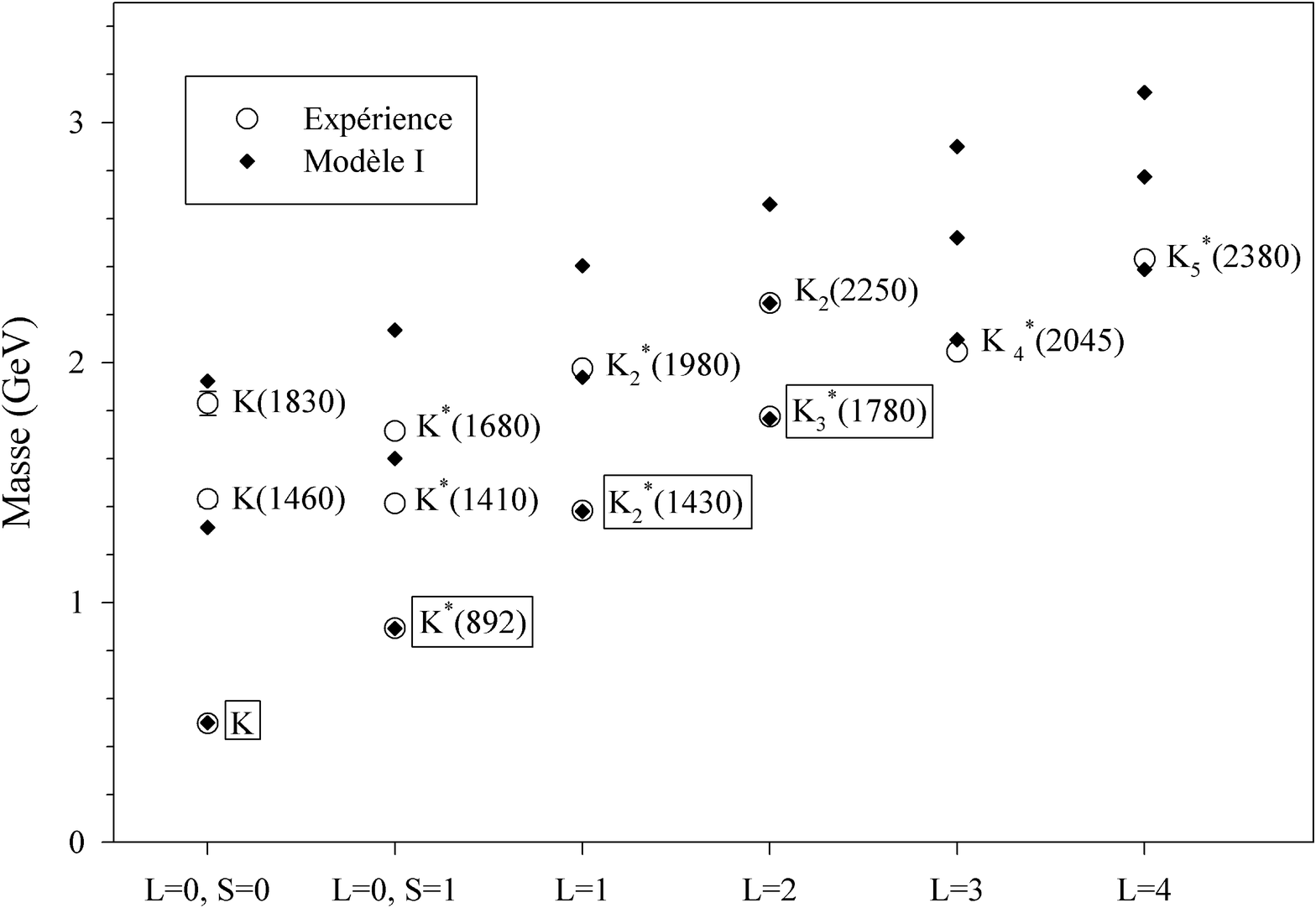}
\protect\caption{Same as for Fig.~\ref{fig:spec1} but for $\bar s n$
mesons.}
\label{fig:spec2}
\end{figure}

\begin{figure}
\centering
\includegraphics*[width=10cm]{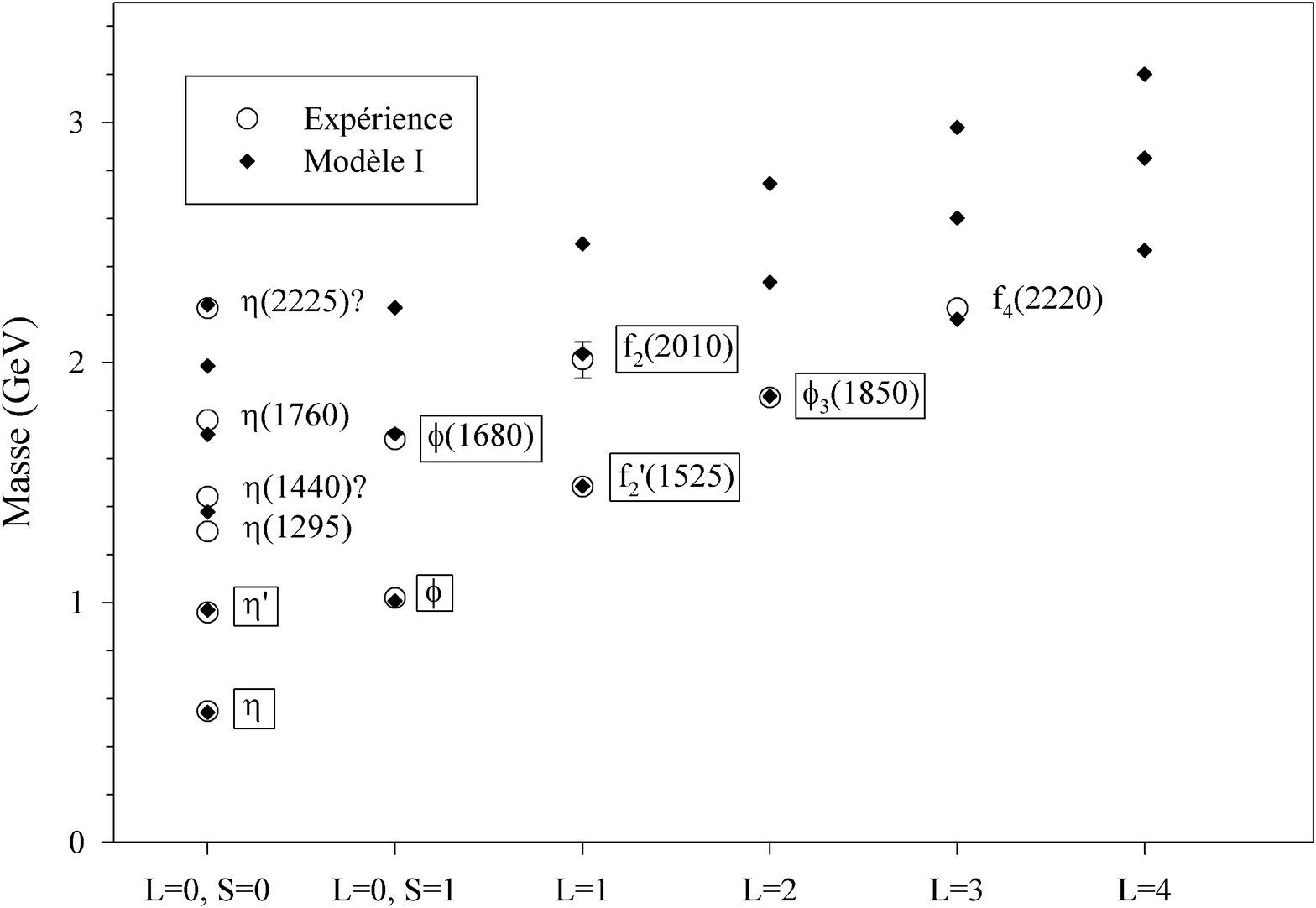}
\protect\caption{Same as for Fig.~\ref{fig:spec1} but for $s \bar s$ and
mixed-flavor mesons.}
\label{fig:spec3}
\end{figure}

 In Figs.~\ref{fig:spec1} to \ref{fig:spec3}, we present the meson
spectra of model~I. We consider this model as the best we have obtained.
Model~III is characterized by a lowest $\chi^2$ value but, as mentioned
above, the corresponding constituent quark masses are too small to hope
to obtain good baryon spectra. The $\chi^2$ value of model~II is also
good but the strange quark size seems very small with respect to usual
estimations found in the literature. In order to decide between models~I
and II, we have tested these two potentials in the heavy meson sector.
By fitting the masses and the sizes of the quarks $c$ and $b$ on 10
c.o.g.\ of $c\bar c$, $b\bar b$, $\bar s n$, $c\bar s$ and $\bar b n$
mesons, we have recalculated a $\chi^2$ value for the all 28 c.o.g.\
considered in this paper (18 for light mesons and 10 for heavy
mesons). The results are indicated in
Table~\ref{tab:param}. Clearly, model~I is preferable. The bad value
obtained for model~II is due to an interplay between the small strange
quark size and the low Coulomb strength.

\par The partial $\chi^2$ for each meson (actually, each c.o.g.\ of
a meson
multiplet) of model~I is below 1.6, except for the $\pi(1300)$-,
$\rho(1450)$- and $a_4(2040)$-mesons. In Fig.~\ref{fig:spec1}, we can
see that the $n\bar n$ mesons are quite well reproduced. The larger
error is obtained for the $\pi$(1300), but the uncertainty on this meson
is very large. We can remark slight deviation from the linear Regge
behavior, but it can be very well reproduced within semi-relativistic
kinematics when only a linear potential is used. 

\par In the $\bar{s}n$ sector, vibrational excitations of the $L=0$
mesons are not satisfactorily reproduced (these states are not taken
into account in the minimization procedure). From the experimental point
of view, the situation is not clear \cite{pdg}. For instance, internal
quantum numbers of $K^*(1410)$ and of $K^*(1680)$ are not well defined
yet. This problem was also revealed in some previous works
\cite{sema95,silv97}. But all the ground states are quite well
reproduced.

\par The $s\bar{s}$ sector is poor in experimental data but all the
states obtained in our model are in quite good agreement with these
data. In the mixed flavor mesons, we reproduce the $\eta$ and $\eta'$
states. The two vibrational excitations of these mesons can be
identified quite well with the $\eta(1295)$ and the $\eta(1760)$. Note
that a calculated state lies between the $\eta(1295)$ and $\eta(1440)$,
which is a non-$q\bar q$ meson candidate \cite{pdg}. At last, we find a
supplementary state between the $\eta(1760)$ and the $\eta(2225)$, the
last one being a not-well established state. 

\subsection{Electromagnetic corrections}
\label{ssec:emc}

The electromagnetic mass differences between mesons are usually ignored
in studies of meson spectra. The reason is that these mass differences
are very small, of the order of some MeV, compared with the orbital and
vibrational excitations due to the strong interaction, which can amount
about one GeV. Nevertheless, it is interesting to calculate the
electromagnetic splittings as they provide further tests on the model.
In particular, the mass differences are very sensitive to the short
range part of the wave functions. 

\par The electromagnetic mass differences between mesons are due to two
distinct effects. A first contribution is provided by the mass
difference between the $u$ and $d$ quarks. In our models we have assumed
that these two quarks have the same mass, but it is no longer relevant
to calculate electromagnetic phenomena. In the following, we will assume
that $m_n = (\bar m_d + \bar m_u)/2$, where $\bar m_i$ is the real
constituent quark mass. We define $\epsilon_d=\bar m_d-m_n$ and
$\epsilon_u=\bar m_u-m_n$. For the strange quark, we have $\bar m_s =
m_s$ and $\epsilon_s=0$. The second contribution is due to the
electromagnetic potential existing between quarks. In first order
approximation, this interaction has the following form
\begin{equation}
\label{vcem}
V_{\text{em}}(r) = \alpha \frac{Q_i Q_j}{r},
\end{equation}
where $\alpha$ is the electromagnetic fine-structure constant, and $Q_i$
is the quark charge for the flavor $i$ in unit of $e$. This
approximation is too crude since the first relativistic corrections of
$V_{\text{em}}$ and $V_{\text{em}}$ have
similar contributions to the electromagnetic mass differences
\cite{luch91}.
We will consider only S-wave mesons, that is to say that the spin-spin
interaction is the only non-vanishing correction. As we work in the
framework of semi-relativistic models, we will use a relativized version
of the spin-spin potential. A method to obtain such a potential is
suggested in Ref.~\cite[p.~198]{luch91} and has been used in
Ref.~\cite{godf85}. The idea is to replace the factors $1/m_i$ by the
operators $1/\sqrt{\vec p\,^2 + m_i^2}$ in the interaction expression. 

\par The contribution of the electromagnetic Hamiltonian $H_{\text{em}}$
can be
calculated within the first order in perturbation theory \cite{silv}.
Assuming that this Hamiltonian is the difference between the total
Hamiltonian (strong plus electromagnetic) and the Hamiltonian used in
our models, the electromagnetic contribution is in first order
approximation
\begin{eqnarray}
\label{hem}
\langle H_{\text{em}} \rangle &=& 
m_i\,\epsilon_i \left\langle \frac{1}{\sqrt{\vec p\,^2 + m_i^2}}
\right\rangle
+ m_j\,\epsilon_j \left\langle \frac{1}{\sqrt{\vec p\,^2 + m_j^2}}
\right\rangle \nonumber \\
&&+ \alpha \left\langle \frac{Q_i Q_j}{r} \right\rangle
- \alpha \frac{8\pi}{3} \left\langle Q_i Q_j \,\vec S_i\cdot \vec S_j 
\left[ \frac{1}{\sqrt{\vec p\,^2 + m_i^2}} \delta(\vec r\,)
\frac{1}{\sqrt{\vec p\,^2 + m_j^2}}\right] \right\rangle,  
\end{eqnarray}
where the symbol $[O]$ is used to indicate the symmetrization of the
operator $O$. The expression~(\ref{hem}) reduces to the usual
non-relativistic form when quark masses tend toward infinity
\cite[p.~169]{luch91}. Mean values of operators can be evaluated as well
by using the harmonic oscillator development of the wave functions as
well directly in configuration space. The isospin breaking between the
$u$ and $d$ quarks is a free parameter in our models, so this quantity
has been fitted to reproduce the $K^+$--$K^0$ mass difference. The
results for model~I are shown in table~\ref{tab:emc}. Other models give
similar results since in all cases the pseudoscalar states are well
described. In magnitude, all results are around two times the
experimental data, but the hierarchy of mass differences is preserved.
In particular the quantity $m_d-m_u$ has the good sign. This is not
always obvious to obtain in potential models \cite{silv98}. It is worth
noting that our models are optimized to reproduce meson mass spectra
with an accuracy of around 10 MeV, and not to obtain the best possible
wave function. So, we consider that the results found are quite well
satisfactory. 

\par We also tried to calculate the electromagnetic mass differences
with the non-relativistic equivalent of Eq.~(\ref{hem}), but the results
obtained are very bad. For instance, isospin breaking between $d$ and
$u$ quark can be found as large as 50 MeV. We have found that the
non-relativistic form of the electromagnetic spin-spin interaction is
responsible of so poor results. This indicates that relativized forms of
operators are important to obtain coherent results in semi-relativistic
models. Since the Coulomb-like strong interaction is of pure vector
type, a strong spin-spin interaction must be introduced if one wants to
take into account the relativistic corrections of the strong potential.
We can guess that the use of a relativized form will be highly
preferable than the usual non-relativistic form. 

\section{Concluding remarks}
\label{sec:concl}

It is well known that the spinless Salpeter equation supplemented by the
so-called funnel potential can describe quite well the main features of
the meson spectra \cite{fulc94}. The mass differences between members of
a $L$-multiplet are generally small with respect to vibrational and
radial excitations. Except in the pseudoscalar meson sector, the
spinless approximation is then generally good. The instanton induced
forces provide a satisfactory way to describe the structure of light
S-wave mesons, including the annihilation phenomenon.

\par Several works are devoted to the study of meson spectra within
relativistic or semi-relativistic models using instanton effects
\cite{munz94,sema95,silv97}. In these works, the instanton parameters
are simply considered as free parameters to be fitted on data. In our
paper, we try to calculate these parameters from the underlying theory.
Our model relies on the spinless Salpeter equation supplemented by the
usual funnel potential and an instanton induced force. Thirteen
parameters are necessary to completely fix the Hamiltonian. This number
could appear large, but six parameters are strongly constrained by
experimental or theoretical considerations, namely, the current quark
masses, the QCD scale parameter, the quark condensates and the string
tension. The Coulomb strength and the constant potential are unavoidable
parameters of potential models. Actually, the Coulomb-like potential
could be replaced by an interaction taking into account the asymptotic
freedom, but it is a complication especially necessary when describing
heavy meson spectra. 

\par The relevance of the remaining five parameters is most
questionable. If quarks are considered as effective degrees of freedom,
then a size as well as a constituent mass can be associated with a
quark. Generally, the quark size is described with a two parameter
function of the constituent mass. As we consider only two different
flavors, it is useless to introduce such a quark size parameterization.
Up to our knowledge, there is no reliable estimation of the quark size,
but our values are in good agreement with values found in other works
\cite{godf85,silv97}. The radial part of the instanton induced potential
is a Dirac-distribution, so it is necessary to replace this form by a
short range potential peaked at origin. The introduction of quark sizes
offers a natural way to realize that, without introducing a new free
parameter which is the instanton range.

\par The two dimensioned coupling constants of the instanton induced
force are calculated from the instanton theory. This interaction
generates also constituent masses for the light quarks. Our work shows
that other contributions must be added to explain the large constituent
masses of
$n$ and $s$ quarks. These contributions which lay between 100 and 200
MeV
are quite large. It is possible to cancel them, but the price to pay is
then to obtain unrealistic very small constituent masses. We consider
that it is better to keep the large values of these contributions. Their
origin is not clear but sea-quark or relativistic effects could explain
their presence.

\par The parameter $\epsilon$ reflects the uncertainties about the form
of the instanton density, which is known within the two-loop
approximation. The values found in all models are consistent with the
fact that the cutoff radius of the instanton size must be small. 

\par We have shown that the light meson spectra can be described by a
semi-relativistic model including an instanton induced force whose
characteristics are calculated from the underlying theory.
Moreover, calculation of electromagnetic mass differences between mesons
indicate that the wave function obtained are quite satisfactory. The
next steps are to calculate meson widths and to generalize the model to
baryons. Such a work is in progress.

\acknowledgments

We thank Professor R. Ceuleneer for useful discussions and constant
interest.
            

\begin{table}
\squeezetable
\protect\caption{Centers of gravity (c.o.g.)\ of $L$- and $I$-multiplets
in GeV for mesons chosen to fix the parameters of the
models. The values of the c.o.g.\ and their corresponding errors are
given by formula~(\ref{cog}). The symbol ``mf '' means ``mixed flavor''.
A meson name used to represent a multiplet in
Figs.~\protect\ref{fig:spec1},
\protect\ref{fig:spec2}, and \protect\ref{fig:spec3} is underlined.}
\label{tab:meson}
\begin{tabular}{lccccc}
State & Flavor & $I$ & $J^{P(C)}$ & $N\ ^{2S+1}L_J$ & c.o.g. \\ 
\hline
$\underline{\pi}$           & $n\bar{n}$ & 1   & $0^{-+}$ & $1\
^{1}S_{0}$ &
0.138$\pm$0.003 \\
\\  
$\omega$                    & $n\bar{n}$ & 0   & $1^{--}$ & $1\
^{3}S_{1}$ &
0.772$\pm$0.001 \\  
$\underline{\rho}$          & $n\bar{n}$ & 1   & $1^{--}$ & $1\
^{3}S_{1}$ &
\\  
\\
$h_1(1170)$                 & $n\bar{n}$ & 0   & $1^{+-}$ & $1\
^{1}P_{1}$ &
1.265$\pm$0.013 \\
$b_1(1235)$                 & $n\bar{n}$ & 1   & $1^{+-}$ & $1\
^{1}P_{1}$ &
\\
$f_1(1285)$                 & $n\bar{n}$ & 0   & $1^{++}$ & $1\
^{3}P_{1}$ &
\\
$a_1(1260)$                 & $n\bar{n}$ & 1   & $1^{++}$ & $1\
^{3}P_{1}$ &
\\
$f_2(1270)$                 & $n\bar{n}$ & 0   & $2^{++}$ & $1\
^{3}P_{2}$ &
\\
$\underline{a_2(1320)}$     & $n\bar{n}$ & 1   & $2^{++}$ & $1\
^{3}P_{2}$ &
\\
\\
$\pi_2(1670)$               & $n\bar{n}$ & 1   & $2^{-+}$ & $1\
^{1}D_{2}$ &
1.681$\pm$0.012 \\
$\omega(1600)$              & $n\bar{n}$ & 0   & $1^{--}$ & $1\
^{3}D_{1}$ &
\\
$\rho(1700)$                & $n\bar{n}$ & 1   & $1^{--}$ & $1\
^{3}D_{1}$ &
\\
$\omega_3(1670)$            & $n\bar{n}$ & 0   & $3^{--}$ & $1\
^{3}D_{3}$ &
\\
$\underline{\rho_3(1690)}$  & $n\bar{n}$ & 1   & $3^{--}$ & $1\
^{3}D_{3}$ &
\\
\\
$f_4(2050)$                 & $n\bar{n}$ & 0   & $4^{++}$ & $1\
^{3}F_{4}$ &
2.039$\pm$0.022 \\
$\underline{a_4(2040)}$     & $n\bar{n}$ & 1   & $4^{++}$ & $1\
^{3}F_{4}$ &
\\
\\
$\underline{\pi(1300)}$     & $n\bar{n}$ & 1   & $0^{-+}$ & $2\
^{1}S_{0}$ &
1.300$\pm$0.100 \\  
\\
$\omega(1420)$              & $n\bar{n}$ & 0   & $1^{--}$ & $2\
^{3}S_{1}$ &
1.454$\pm$0.026 \\  
$\underline{\rho(1450)}$    & $n\bar{n}$ & 1   & $1^{--}$ & $2\
^{3}S_{1}$ &
\\  
\\
$\underline{K}$             & $\bar{s}n$ & 1/2 & $0^{-}$ & $1\
^{1}S_{0}$ &
0.496$\pm$0.002 \\
\\
$\underline{K^*(892)}$      & $\bar{s}n$ & 1/2 & $1^{-}$ & $1\
^{3}S_{1}$ &
0.892$\pm$0.001 \\
\\
$K_1(1270)$                 & $\bar{s}n$ & 1/2 & $1^{+}$ & $1\
^{1}P_{1}$ &
1.382$\pm$0.005 \\
$K_0^*(1430)$               & $\bar{s}n$ & 1/2 & $0^{+}$ & $1\
^{3}P_{0}$ &
\\
$K_1(1400)$                 & $\bar{s}n$ & 1/2 & $1^{+}$ & $1\
^{3}P_{1}$ &
\\
$\underline{K_2^*(1430)}$   & $\bar{s}n$ & 1/2 & $2^{+}$ & $1\
^{3}P_{2}$ &
\\
\\
$K_2(1770)$                 & $\bar{s}n$ & 1/2 & $2^{-}$ & $1\
^{1}D_{2}$ &
1.774$\pm$0.012 \\
$K^*(1680)$                 & $\bar{s}n$ & 1/2 & $1^{-}$ & $1\
^{3}D_{1}$ &
\\
$K_2(1820)$                 & $\bar{s}n$ & 1/2 & $2^{-}$ & $1\
^{3}D_{2}$ &
\\
$\underline{K_3^*(1780)}$   & $\bar{s}n$ & 1/2 & $3^{-}$ & $1\
^{3}D_{3}$ &
\\
\\
$\underline{\phi}$          & $s\bar{s}$ & 0   & $1^{--}$ & $1\
^{3}S_{1}$ &
1.019$\pm$0.001 \\
\\
$h_1(1380)$                 & $s\bar{s}$ & 0   & $1^{+-}$ & $1\
^{1}P_{1}$ &
1.482$\pm$0.009 \\
$f_1(1510)$                 & $s\bar{s}$ & 0   & $1^{++}$ & $1\
^{3}P_{1}$ &
\\
$\underline{f_2'(1525)}$    & $s\bar{s}$ & 0   & $2^{++}$ & $1\
^{3}P_{2}$ &
\\
\\
$\underline{\phi_3(1850)}$  & $s\bar{s}$ & 0   & $3^{--}$ & $1\
^{3}D_{3}$ &
1.854$\pm$0.007 \\
\\
$\underline{\phi(1680)}$    & $s\bar{s}$ & 0   & $1^{--}$ & $2\
^{3}S_{1}$ &
1.680$\pm$0.020 \\
\\
$\underline{f_2(2010)}$     & $s\bar{s}$ & 0   & $2^{++}$ & $2\
^{3}P_{2}$ &
2.011$\pm$0.080 \\
\\
$\underline{\eta}$          & mf       & 0   & $0^{-+}$ & $1\ ^{1}S_{0}$
&
0.547$\pm$0.001 \\
$\underline{\eta'}$         & mf       & 0   & $0^{-+}$ & $1\ ^{1}S_{0}$
&
0.958$\pm$0.001 
\end{tabular}
\end{table}

\begin{table}
\squeezetable
\protect\caption{Optimal parameters found for various models (see
Sec.~\protect\ref{ssec:spectra}). Values between square brackets are
fixed quantities in the model. When available, the expected value of a
parameter is also given in the column ``Exp.''. The {\em Ans\"{a}tze} to
calculate $\gamma_{\text{mf}}$ (see Sec.~\protect\ref{ssec:eff}) is
indicated and the parameters $m_n$, $m_s$, $g$, $g'$ and
$\rho_{\text{c}}$ are calculated. The quantity $\chi^2$(light) is the
value of the $\chi^2$ given by relation~(\protect\ref{chi2}) for the set
of mesons from Table~\protect\ref{tab:meson}. The corresponding quantity
for an extended set of mesons (see Sec.~\protect\ref{ssec:spectra}) is
indicated by $\chi^2$(all).}
\label{tab:param}
\begin{tabular}{rlcccccc}
\multicolumn{2}{c}{
Parameter} & Model I & Model II & Model III & Model IV & Model V & Exp.
\\ 
\hline
$m^{0}_n$         & (GeV)       & 0.015   & 0.013   & 0.004   & 0.015 &
0.015 & 0.002--0.015\protect\cite{pdg}  \\
$m^{0}_s$         & (GeV)       & 0.215   & 0.196   & 0.199   & 0.271 &
0.166 & 0.100--0.300\protect\cite{pdg}  \\
$\Lambda$  & (GeV)       & 0.245   & 0.299   & 0.223   & 0.231 &
0.262 & 0.209$^{+0.039}_{-0.033}$\protect\cite{pdg} \\
$\langle\bar{n}n\rangle$       & (GeV$^3$)       & ($-$0.243)$^3$
& ($-$0.223)$^3$ & ($-$0.236)$^3$ & ($-$0.250)$^3$  
& ($-$0.240)$^3$ & $(-0.225\pm 0.025)^3$\protect\cite{rein85}  \\
$\langle\bar{s}s\rangle/\langle\bar{n}n\rangle$    &             & 0.706
&
0.834   & 0.704   
& 0.700 & 0.703 &  $0.8 \pm 0.1$\cite{rein85} \\
$a$              & (GeV$^2$)   & 0.212   & 0.225   & 0.207   & 0.201 &
0.235
& $0.20 \pm 0.03$\cite{mich96}  \\
$\kappa$         &             & 0.440   & 0.283   & 0.366   & 0.367 &
[0.000] &   \\
$C$              & (GeV)       &$-$0.666 &$-$0.781 &$-$0.593 &$-$0.551 &
$-$0.860 &   \\ 
$\gamma_{n}$     & (GeV$^{-1}$)& 0.736   & 0.730   & 0.764   & 0.773 &
0.586
&   \\
$\gamma_{s}$     & (GeV$^{-1}$)& 0.515   & 0.186   & 0.552   & 0.524 &
0.409
&   \\
$\delta_{n}$     & (GeV)       & 0.120   & 0.147   & [0.000]   & 
[0.000] & 0.120 &   \\
$\delta_{s}$     & (GeV)       & 0.173   & 0.228   & 0.118   &
[0.000] & 
0.194 &   \\
$\epsilon$ &  & 0.031 & 0.162 & 0.011 & 0.046 & 0.003 & \\
$\gamma_{\text{mf}}$ & & $\sqrt{2\gamma_n\gamma_s}$ & 
$\sqrt{\gamma_n^2 + \gamma_s^2}$
 & $\sqrt{2\gamma_n\gamma_s}$ &
$\sqrt{2\gamma_n\gamma_s}$ & $\sqrt{2\gamma_n\gamma_s}$ & \\
\hline
$m_n$            & (GeV)       & 0.192   & 0.206   & 0.061   & 0.090 &
0.176
&   \\
$m_s$            & (GeV)       & 0.420   & 0.443   & 0.351   & 0.312 &
0.385
&   \\
$g$              & (GeV$^{-2}$)& 2.743   & 2.767   & 3.117   & 3.336 &
2.002
&   \\
$g'$             & (GeV$^{-2}$)& 1.571   & 1.213   & 1.846   & 1.807 &
1.241
&   \\
$\rho_{\text{c}}$         & (GeV$^{-1}$)& 1.541   & 1.402   & 1.664   &
1.598
& 1.461 &   \\
\hline
\multicolumn{2}{c}{$\chi^2$(light)}              
& \multicolumn{1}{r}{14.9}    &
\multicolumn{1}{r}{16.7}  & 
\multicolumn{1}{r}{10.9}    & \multicolumn{1}{r}{73.1} &
\multicolumn{1}{r}{43.9}
&   \\
\multicolumn{2}{c}{$\chi^2$(all)}
                & \multicolumn{1}{r}{35}    & \multicolumn{1}{r}{95.5}
                &
                &  &       &    
\end{tabular}
\end{table}

\begin{table}
\protect\caption{Electromagnetic mass differences for some mesons in
MeV. 
The results of our model~I is compared with the data.}
\label{tab:emc}
\begin{tabular}{lrr}
 & Model I & Experiment\protect\cite{pdg} \\
\hline
$m_{\pi^+}-m_{\pi^0}$ & 7.184 & $4.594\pm 0.001$ \\
$m_{\rho^+}-m_{\rho^0}$ & 1.361 & $-0.300\pm 2.200$ \\
$m_{K^+}-m_{K^0}$ & fitted & $-3.995\pm 0.034$ \\
$m_{K^{*+}}-m_{K^{*0}}$ & $-$11.047 & $-6.700\pm 1.200$ \\
\hline
$m_d - m_u$ & 26.610
\tablenote{
fitted to reproduce the $m_{K^{+}}-m_{K^{0}}$ difference.} &
$\lesssim 15$ 
\end{tabular}
\end{table}

\end{document}